\documentclass[12pt, preprint]{aastex}

\usepackage{graphicx}
\shorttitle{Streamer-Shock Type II}

\shortauthors{Kong et al.}

\usepackage{color}
\usepackage[normalem]{ulem}

\usepackage{multirow}

\begin{document}

\title{Possible role of coronal streamer as magnetically-closed structure in shock-induced energetic electrons and metric type II radio bursts}

\author{Xiangliang Kong\altaffilmark{1},
Yao Chen\altaffilmark{1},  Fan Guo\altaffilmark{2}, Shiwei
Feng\altaffilmark{1},  Bing Wang\altaffilmark{1}, Guohui
Du\altaffilmark{1}, and  Gang Li\altaffilmark{3}}

\altaffiltext{1}{Shandong Provincial Key Laboratory of Optical Astronomy and Solar-Terrestrial Environment,
and Institute of Space Sciences, Shandong University, Weihai, Shandong 264209,
China; yaochen@sdu.edu.cn} \altaffiltext{2}{Theoretical Division,
Los Alamos National Laboratory, Los Alamos, NM 87545, USA}
\altaffiltext{3}{Department of Space Science and CSPAR, University
of Alabama in Huntsville, Huntsville, AL 35899, USA}

\begin{abstract}

Two solar type II radio bursts, separated by $\sim$24 hours in
time, are examined together. Both events are associated
with coronal mass ejections (CMEs) erupting from the same active
region (NOAA 11176) beneath a well-observed helmet streamer.
We find that the type II emissions in both events ended once the
CME/shock fronts passed the white-light streamer tip, which is
presumably the magnetic cusp of the streamer. This leads us to conjecture that
the closed magnetic arcades of the streamer may play a role in electron acceleration and type
II excitation at coronal shocks. To examine such a conjecture, we
conduct a test-particle simulation for electron dynamics within a
large-scale partially-closed streamer magnetic configuration swept
by a coronal shock. We find that the closed field lines play the role of an electron trap,
via which the electrons are sent back to the shock front for multiple times,
and therefore accelerated to high energies by the shock.
Electrons with an initial energy of $300$ eV can be accelerated to tens of keV
concentrating at the loop apex close to the shock front with a counter-streaming distribution at most locations.
These electrons are energetic enough to excite Langmuir
waves and radio bursts. Considering the fact that most solar eruptions originate from closed field
regions, we suggest that the scenario may be important to the generation of more metric type IIs.
This study also provides an explanation to the general ending frequencies of metric type IIs
at or above 20-30 MHz and the disconnection issue between metric and
interplanetary type IIs.

\end{abstract}


\section{Introduction}
Streamers are the most prominent quasi-steady structures in the
solar corona. Coronal Mass Ejections (CMEs) are observed to be
closely related to and frequently interact with helmet streamers
\citep[e.g.,][]{howard85,hundhausen93,mcallister96,chen10,chen11,feng11,chen13}.
In the streamer region, the Alfv\'enic speed is lower than that of
the surroundings because of a much higher plasma density, and the
plasma outflow is slow and hardly measurable in the closed
magnetic field region \citep{habbal97,strachan02,kwon13,chen13}.
Due to these special plasma and magnetic conditions, the streamer
structure is expected to facilitate the formation and/or
enhancement of a CME-driven shock.

In recent studies, streamers are found to be important on the
generation of type II radio bursts and the morphology of radio
dynamic spectra. For example, it is suspected that the interaction
region between CME/shock flanks and streamers is an important
source of type II radio bursts
\citep{reiner03,mancuso04,cho07,cho08,cho11,feng12,feng13,shen13,chen14,magdalenic14}.
Lately, \citet{kong12} and \citet{feng12,feng13} studied the
effect of dense streamer structures on the type II spectra, and
further inferred properties of the type II sources. According to
the plasma emission mechanism, the type II radiation frequency is
mainly determined by the local electron plasma density
\citep{ginzburg58,nelson85}. Therefore, the type II spectral shape
may change accordingly if dense structures (e.g., streamers) are
present in the path of type II emitting sources. Two specific
morphological features of type IIs relevant to CME-streamer
interactions, namely, the spectral break and spectral bump, were
identified. The spectral break appears as an abrupt change of the drifting rate of the type II spectrum
while the bump appears as a plateau or bulge on the type II spectrum.
\citet{feng13} also suggested that the type II source
should be spatially compact with a spatial size less than 0.05-0.1
$R_\odot$, according to the overall bumping time of the emission
lanes. Although the above studies are useful in unravelling the
connection between type II bursts and CME-streamer interaction,
 how streamers are physically involved in electron
acceleration and radio emission processes induced by a coronal
shock remains unknown.

The observed type II bursts indicate the generation of nonthermal
electrons close to coronal shocks \citep{nelson85}. Previous
studies often consider the acceleration of electrons at a planar
shock \citep[e.g.,][]{holman83,wu84}. However, it is known that
this can only lead to strong acceleration when the shock normal is
quasi-perpendicular to the incident magnetic field and the shock
speed is high. Recent works have investigated nonplanar effects on the
acceleration of particles. For instance, magnetic trapping geometries have
been considered to be efficient particle accelerator when there
exists large-scale magnetic turbulence interacting with a shock
\citep[e.g.,][]{guo10,guo10etal}, at the termination shock induced
by a solar flare reconnection downflow
\citep[e.g.,][]{somov97,guo12,nishizuka13}, and in the magnetotail
dipolarization process during magnetic storms or substorms in the
absence of a shock \citep[e.g.,][]{birn97,birn98}.
As a CME-driven shock sweeps through a streamer, the shock front can
intersect with the same magnetic field line at two different
points. This leads to a similar magnetic trapping geometry likely important
to shock-induced electron acceleration and type II excitation.

In this study, we explore the possible role of streamers as magnetically-closed structures in electron
acceleration and excitation of metric type II radio bursts induced by a coronal shock from the following
two aspects. We first present observations of two type II burst
events to indicate that the closed structure of coronal streamer may be important to type II excitations.
We then conduct a test-particle simulation to demonstrate that the shock-streamer system can
lead to strong electron acceleration, validating the physical implication derived from observations.
Conclusions and discussion are presented in the last section.

\section{Observation of two metric type II radio bursts}\label{sec2}

In this section, we analyze the observational data of two metric
type II radio bursts that occurred on 2011 March 25 and 27, with a
$\sim$24-hour separation. The later event has been studied in
detail by \citet{kong12}. Here we first focus on the former event
by examining the simultaneous radio spectra and CME imaging data.
We then briefly summarize common observational features of the two
events.

Figure 1 shows the radio dynamic spectrum between 23:13 UT and
23:39 UT on 2011 March 25. The data are a combination of Bruny
Island Radio Spectrometer \citep[BIRS;][]{erickson97} in the range
of 10-35 MHz and Learmonth in the range of 35-180 MHz. The
temporal resolution of both stations is 3 seconds. Other radio
stations, such as Culgoora \citep{prestage94} and Green Bank Solar
Radio Burst Spectrometer \citep[GBSRBS;][]{white06}, also recorded
this radio burst. Before the onset of the type II burst, we
observe a fast-drifting type III radio burst at $\sim$23:15 UT
during the impulsive phase of an M1.0 GOES soft X-ray flare. The
type II burst starts at $\sim$23:17 UT. Both the fundamental and
harmonic bands can be identified, as denoted by ``F'' and ``H'' in
Figure 1. From the harmonic band, we can obtain the average
 drift rate of the type II burst being $\sim$$-$0.18 MHz
s$^{-1}$ ($\sim$$-$0.09 MHz s$^{-1}$ for the fundamental band),
and the ending time is $\sim$23:36 UT. We can see some bump- or
break- like morphological features of the slowly-drifting
backbone. They are probably caused by the shock penetrating into
or propagating across coronal density structures
\citep{feng12,feng13,kong12}.

Now we examine imaging observations of the eruption process. The
CMEs in these two events erupted from the same active region (AR)
NOAA 11176. By combining the imaging data observed by the Solar
Dynamics Observatory (SDO)/AIA \citep{lemen12} and the Solar
TErrestrial RElations Observatory \citep[STEREO;][]{kaiser08} and
the magnetic field configuration obtained from the potential-field
source-surface model \citep[PFSS;][]{schatten69,schrijver03},
\citet{kong12} concluded that the AR was located at one foot of a
large-scale streamer structure. On 2011 March 25, the location of
the AR was S16E30, and STEREO A (B) was $\sim$89$^{\circ}$ ahead
($\sim$95$^{\circ}$ behind) of the Earth. Therefore, in the field
of view (FOV) of the Earth the eruption is a solar disk event,
while in the FOV of STEREO A a backside event near the east limb
and in the FOV of STEREO B a front side event near the west limb.
This greatly reduces the projection effect on measuring the CME front heights with STEREO.

Figure 2 shows the white light
coronagraph data observed by STEREO/COR1 B between 23:20 UT and
23:55 UT. Panel (a) is the direct image at 23:20 UT and panels
(b)-(h) are the running difference images. From these images we
can see that an outward-propagating bright CME front moves through
the streamer structure.
The CME front becomes much weaker and more diffuse when passing over the main structure of the streamer.
The blue triangles are the outlining streamer envelop depicted from panel (a) and over-plotted onto other panels,
to indicate the relative location of the streamer and the CME front.

Heliocentric heights of the CME front
measured along the streamer axis are shown as squares in the height-time plot in Figure 3,
with blue for COR1 A and red for COR1 B. The measurement
uncertainty of the heights from COR1 is estimated to be less than
0.05 $R_\odot$ ($\sim$5 pixels). From the linear fit to the data
points, we find that the mean speeds of the CME front at
23:25-23:40 UT are 514 $\pm$ 13 km s$^{-1}$ (COR1 A) and 520 $\pm$
29 km s$^{-1}$ (COR1 B). At the time of the eruption, the temporal
cadence of EUVI B in the 195 \AA\ wavelength is 2.5 min, while it
is 5 min for EUVI A. As noted above, the AR is on the backside as
seen from STEREO A, therefore we only show the heights of the
eruption front measured from EUVI B, as red triangles. The
measurement uncertainty is estimated to be about 0.02 $R_\odot$
($\sim$10 pixels). Using the data points of both EUVI B and COR1
B, we can obtain the mean speed of the CME front during the type
II burst (23:17-23:36 UT) to be $\sim$620 $\pm$ 25 km s$^{-1}$. In
addition, in Figure 3 we also plot the heights of coronal EUV
wave observed by AIA in 193 \AA\ (taken from \citet{kumar13}) with
green plus signs.

With the mean speed of CME front ($\sim$ 620 km s$^{-1}$) and the
two-fold Newkirk density model \citep{newkirk61}, the type II
spectrum can be fitted well as shown by the red solid line in Figure
1. To illustrate the influence of using different density models,
we also present the fitting curves using one-fold (red dashed
line) and three-fold (red dotted line) Newkirk model. It can be
seen that these three lines can basically enclose the entire type
II emissions. The heights of radio emission source deduced from
the spectral fitting are shown in Figure 3 as three black
straight lines. The dashed (solid, dotted) line represents the
fitting curve with the 1-fold (2-fold, 3-fold) Newkirk density
model. Comparing the heights of the coronal eruption front
measured from the imaging observations by SDO and STEREO to the
radio source heights obtained by the spectral fitting, we find
that different data sets of heights are consistent with each
other. On the other hand, from panels (c)-(h), there exits an apparent deflection of coronal ray/streamer on the southern flank of the CME corresponding to its lateral expansion, which indicates a fast-wave nature of the CME front. Based on these observations, we suggest that the type II
burst was generated by the eruption-driven front (presumably the shock) from inside of the streamer.
Such a suggestion is supported by many previous studies which provide important insights into the
physical relation between the type-II-emitting shock and the EUV/white-light front
(e.g., Biesecker et al. 2002; Vrsnak et al. 2006; Chen et al. 2014).

For the purpose of our study, it is important to determine the height of streamer cusp. The height can be estimated from the direct white-light image of the streamer as shown in Figure 2(a). The top of the outlining envelope can be used to determine the cusp location. To provide a consistency check, we examine carefully the difference images shown in Figure 2. From panels (e)-(g), we observe a rising bright cusp feature within the streamer envelope. The cusp stops rising after 23:50 UT till it becomes overlapping with the streamer envelope, and remains bright till 00:20 UT the next day. In the meanwhile, the CME front continues its outward propagation. The bright cusp-like difference structure is formed by the upwelling plasmas along the disturbed yet still-closed (or just-closed) streamer structure in the recovering phase of the post-eruption corona. These plasmas are contained by the closed field lines and accumulated there, as indicated by the observation that the cusp structure stops rising yet still remains bright after it coincides with the white-light streamer envelope. We therefore suggest that the cusp feature observed in panels (g) and (h) can be used to depict the closed streamer structure and to determine its cusp height. In addition, the same feature also appears in the base difference image (Figure 2(i)), providing a consistency check of our analysis. With this analysis, the cusp height is estimated to be at 2.2 - 2.4 $R_\odot$.
Similar cusp structure also appears both in the running difference and base difference images for the 2011 March 27 event (e.g., see Figure 4(f) in Kong et al. (2012)). The heights of streamer tip (cusp) are similar in both events.

It can be seen that the CME/shock front
propagated across the streamer cusp region at 23:35-23:40 UT. In
Figure 1 we see that the type II burst ends at $\sim$23:36 UT.
Therefore, within the observational uncertainty, we can infer that
the type II ending time coincides with the time when the shock
passed the streamer cusp.

For the event on 2011 March 27, the type II dynamic spectrum and
the eruption process have been investigated in \citet{kong12}. The
main feature of that event is that the type II spectrum shows an
intriguing break, i.e. the normal slowly-drifting emission being
followed by a few fast-drifting bands. They suggested that the
pre-break emission was produced by the shock propagating within
the streamer, while the spectral break was caused by the
radio-emitting shock crossing the streamer boundary along which
the plasma density drops abruptly.

Some common observational features of the two events are
summarized as follows: (1) Both CMEs erupted from the same AR
beneath a well-observed helmet streamer, and the sweeping process
of the CME front through the streamer structure can be observed
clearly; (2) the heights of CME front obtained from
imaging observations are consistent with that deduced from the
type II spectral fitting using a reasonable density model; (3)
type II radio emission ended when CME/shock fronts passed the
streamer cusp region, subject to observational uncertainty. These
observational results indicate that both type IIs were
possibly related to the shock-streamer interaction. Especially, the last point
leads us to conjecture that the type IIs are likely affected by the magnetically-closed streamer configuration.
To test such an observational indication, we conduct test-particle
simulation of electron acceleration in a streamer-shock system.

\section{Test-particle simulation of electron acceleration in
 a streamer-shock system}\label{sec3}

In this section, we carry out a test-particle simulation to study
the energization of electrons in a prescribed magnetic
configuration consisting of a streamer and an outward-propagating
shock. In Section 3.1, we introduce the initial configuration and
parameter setup of the streamer-shock system and the test-particle
simulation. In Section 3.2, we first analyze the trajectory and
energy evolution of a typical electron to understand the
acceleration mechanism; then we compare the distributions of
electrons that are accelerated to different energies to identify
the factors that affect the electron acceleration process; we also
present the energy spectra of energetic electrons.

\subsection{Numerical Model}

We use an analytical model for streamers given by
\citet{low86}. It describes an axisymmetric magnetic structure
containing both magnetic arcades and open field lines with a
current sheet in a spherical coordinate ($r$, $\theta$). This
model has been used in previous corona and solar wind modellings \citep[e.g.,][]{chen01,hu03a,hu03b}. In
this study, the magnetic field strength in the polar region on the
solar disk is set to be 10 G. The magnetic topology of the
streamer in the region of interest is shown in Figure 4 under
Cartesian coordinate ($x$, $z$). The $z$-axis represents the
rotation axis of the Sun, the $x$-axis is in the solar equatorial
plane parallel to the streamer axis, and the $y$-axis completes
the right-handed triad with the solar center being at the origin.
The black lines represent magnetic field lines and the red line
denotes the outermost closed field line and the current sheet
above. The height of streamer cusp is taken to be 2.5 $R_\odot$.
The simulation domain is given by $x$ = [1.5, 3.0] $R_\odot$ and
$z$ = [-0.8, 0.8] $R_\odot$, which includes some open fields
surrounding the streamer. The $y$ component of the magnetic field
$B_y$ is set to be 0. In the simulation domain, the average
magnitude of magnetic field ($B_0$) is $\sim$0.2 G.

For simplicity, we consider a planar shock propagating along
the streamer axis, as shown by the dashed blue line in Figure 4.
The shock is assumed to form at $x_0$ = 1.5 $R_\odot$
 in the beginning of the simulation, consistent with previous studies of the formation heights
 of metric type II shocks \citep[e.g.,][]{pohjolainen08,magdalenic08,nindos11,gopalswamy09}.
The shock speed $U_{sh}$ is taken to be 600 km s$^{-1}$.
We also assume the shock is wider than the
streamer. The calculation is carried out in the
shock frame ($x'$, $y$, $z$), where the shock is at $x' = 0$,
and the plasmas carrying the magnetic field flow from $x' < 0$
(upstream) with a speed of $U_1 \sim U_{sh}$ to $x' > 0$
(downstream) with a speed of $U_2$. The flow speed close to the
shock is given by a hyperbolic tangent function U($x'$) = ($U_1$ +
$U_2$)/2 $-$ ($U_1$ $-$ $U_2$) tanh($x'$/$th$)/2, where $U_1$ and
$U_2$ are the upstream and downstream flow speeds in the
shock frame. The shock compression ratio is assumed to be
$U_1/U_2=4$ for simplicity. $th$ is the shock
thickness and is taken to be $0.01$ $U_1$/$\Omega_{ci}$, where
$\Omega_{ci}$ is the proton gyrofrequency defined by $B_0$.
Using a larger value of $th$ $\sim$0.1 $U_1$/$\Omega_{ci}$ in the
simulation (not shown here) does not change our
results considerably.

After the shock starts to propagate outward from 1.5 $R_\odot$,
electrons with an initial energy $E_0$ = 300 eV are continuously
injected at a constant rate at $x'$ = $-$10 $U_1/\Omega_{ci}$ in the upstream.
The initial pitch angles of the
electrons are given randomly. For each electron, the equation of
motion under the Lorentz force is solved in the shock frame. The
electron mass is taken to be $1/1836$ of the proton mass. The
numerical technique used to integrate electron trajectories is the
Bulirsch-Stoer method \citep{press86}, which has been widely used
in calculating particle trajectories \citep[e.g.,][]{guo14}. The
algorithm uses an adjustable time-step method based on the
evaluation of the local truncation error. It is highly accurate
and has been tested to conserve particle energy to a very good
degree. In this study, a total of 1.5$\times$10$^6$ electrons are
injected. When an electron moves out of the simulation domain or
reaches a distance of $10^4$ $U_1/\Omega_{ci}$ downstream of the
shock, we stop tracking it and terminate the calculation. An \textit{ad hoc} pitch-angle
scattering is included to mimic the effects of coronal plasma turbulence,
kinetic waves on electron and ion scales, and Coulomb collisions
\citep[e.g.,][]{Marsch2006}. This is done by
randomly changing the electron pitch angle every 
$\tau =$ 10$^4$ $\Omega_{ci}^{-1}$.

\subsection{Simulation Results}

The simulation results show that low energy electrons can be
accelerated to an energy up to several hundred times the
initial energy $E_0$. We first analyze the simulation result for
an electron that is accelerated to $\sim$25 $E_0$ to show the shock acceleration mechanism.

In Figure 5, we display the electron trajectory in the $x$-$z$ lab frame (panel a)
and in the $x'$-$z$ shock frame (panel c). The blue arrows in
these two panels point to the injection point of the electron.
From panel (c) we can see that the electron propagates in the
closed field and interacts with the shock front multiple times.
Panels (b) and (d) show its position $x$ and $x'$ over time,
respectively. The dashed blue line in panel (b) indicates the
position of the outward propagating shock, while that in panel (d)
indicates the shock front ($x'$ = 0) in the shock frame. The
vertical red dotted lines denote the electron reflection points at
the shock. As seen from our simulation results, at the shock an
electron can either get reflected back to the upstream or go
through the shock moving to the downstream. This is mainly
determined by the exact value of the electron pitch angle and the
magnetic field variation induced by the shock. As the shock passes
over a closed field line, the electrons moving along that field
line are left behind in the downstream.

In panels (e)-(f) we present the temporal evolution of the electron drifting distance
along the $y$ direction and the temporal evolution of its energy
in the shock frame. It can be seen that a fast drift in the $y$
direction is accompanied by a simultaneous sharp increase of the
electron energy whenever the electron is reflected at the shock
(see the vertical red dotted lines). Therefore, the electron gains
energy mainly during its gradient-\textbf{B} drift along the
shock-induced electric field (-\textbf{U}$\times$\textbf{B}). In
other words, the electron is accelerated via the well-known shock
drift acceleration (SDA) mechanism (also called the fast Fermi
acceleration at a quasi-perpendicular shock) \citep{armstrong85,
wu84}.

It can be seen from Figures 5(c) and 4(d) that an electron may
also change its direction besides reflection at the shock front
($x'$ = 0). This is due to the pitch angle scattering of
electrons. To examine the effect of such scattering on our result,
we also conduct simulations without any scattering. In the above case
(with a random scattering every $\tau =$ 10$^4$ $\Omega_{ci}^{-1}$),
electrons can be energized up to $\sim$530 $E_0$, and the
fractions of electrons accelerated to $>$5 $E_0$, $>$10 $E_0$,
$>$20 $E_0$ and $>$30 $E_0$ are $\sim$8.4\%, 3.9\%, 1.1\% and
0.39\%, respectively. In comparison, for the case without any
scattering, the highest electron energy obtained is $\sim$48
$E_0$, and the fractions of electrons accelerated to $>$5 $E_0$,
$>$10 $E_0$, $>$20 $E_0$ and $>$30 $E_0$ reduce to $\sim$2.4\%,
0.25\%, 0.005\% and 0.002\%, respectively. This suggests that the
scattering effect plays an important role in the electron acceleration process,
which allows electrons to get more chance to encounter the shock
front and thus receive more accelerations. Recent spacecraft
observations and numerical simulations have shown that whistler
waves and small-scale shock ripples can play a role in scattering
electrons in pitch-angle at the shock front \citep{burgess06,guo10,wilson12}.
Indeed, enhanced scattering implies a diffusive shock acceleration (DSA) process.
Note that although the effect
of pitch angle scattering can affect our quantitative results, it
does not change the main conclusion of this study.

To analyze other physical factors affecting the acceleration, we
examine the distribution of injection position of electrons that
have achieved a final energy of 5-10 $E_0$, 10-20 $E_0$, 20-30
$E_0$ and $>$30 $E_0$. Figure 6 presents this distribution. The
most obvious feature of this figure is that none of the electrons
that are injected in the open field regions are accelerated to
$>$5 $E_0$. In other words, only electrons injected in the closed
field region can be efficiently accelerated by the SDA mechanism of the study. Note,
however, we do not include the effect of perpendicular diffusion
due to, e.g., large-scale magnetic turbulence
\citep[e.g.,][]{giacalone99}. The same conclusion is obtained in
the case without any scattering. This highlights the importance of
the large-scale closed magnetic field in shock-induced electron
acceleration.

In addition, fewer accelerated electrons have an injection
position further away from the streamer axis ($\left|z\right|>$0.4-0.5
$R_\odot$). This may be partly due to the location-dependent shock
geometry. To illustrate this point, in Figure 7 we show variations
of $\theta_{Bn}$ (the angle between the upstream magnetic field
line and the shock normal) with $z$ at different distances ($x$ =
1.5, 1.6, 1.8 and 2.0 $R_\odot$). We find that the shock is more
quasi-perpendicular ($\theta_{Bn} > 45^\circ$) closer to the
streamer axis, and more quasi-parallel ($\theta_{Bn} < 45^\circ$)
further away from the streamer axis. According to earlier studies,
the quasi-perpendicular shock geometry favors electron
acceleration \citep[e.g.,][]{holman83,wu84,guo10}. This is
consistent with our simulation result. Another factor which may contribute
to this is that generally electrons moving along shorter loops
(with injection locations closer to the streamer axis)
need less time to approach the shock front, thus less time is required for further electron reflection and acceleration.
Nevertheless, only few electrons that are injected very close to the streamer axis
($|z|<$0.02-0.03 $R_\odot$) are accelerated, as seen from this
figure. Possible explanations for this result are twofold. First,
these electrons are released right atop of closed field lines and
the shock front is nearly parallel to the upstream field and about
to embrace the field line, so there may not be enough time for
electrons to get repetitive accelerations. Second, as pointed out
by \citet{holman83}, if $\theta_{Bn}$ is very close to $90^\circ$,
the reflection condition can not be satisfied, therefore these
electrons can not receive efficient acceleration.

Now we examine the electron positions at 9 different times
when the shock propagates to distances of 1.6, 1.7, 1.8, ..., 2.4
$R_\odot$. These data sets are superposed onto the
streamer field lines and shown together in Figure 8, where the
scattering points represent electrons that have been accelerated
to 5-10 $E_0$ (panel a), 10-20 $E_0$ (b), 20-30 $E_0$ (c), and
$>$30 $E_0$ (d). It can be seen that energetic electrons mainly
concentrate in the shock upstream within its immediate
neighborhood, and close to the streamer axis (i.e., around the tip
of the relevant closed field lines).
In the above paragraph, we have presented the two possible factors making the electron acceleration
at the loop apex (close to the streamer axis) more efficient.

In addition, electrons of higher energies are more concentrated in a narrower region.
Electrons shown in panels (c) and (d) have energy of $>$20-30
$E_0$ ($>$6-9 keV with an electron velocity of $>$0.15-0.2 $c$,
$c$ is the speed of light). Previous studies suggest that electron
beams with such high energies are capable of exciting Langmuir
waves and radio emission \citep[e.g.,][]{ergun98,mann05}. Thus, it
is reasonable to regard that the region with $>$20-30 $E_0$
electrons is the likely source of the radio bursts.
As seen from panels (c) and (d), the region with energetic electrons
(presumed to be the radio source region) concentrate around the
top of the closed field lines with a radial extension of only a few hundredths $R_\odot$ and
a latitudinal extension of a few tenths $R_\odot$. In other words, according our simulation, the outermost
part of a closed field line is filled with electrons that are
energetic enough to excite Langmuir oscillations and may become
radio loud while the field line is about to be embraced by the
shock. This simulation result provides an explanation to the well-known fact that the type II emission is generally
confined to fairly narrow frequency bands as shown in Figure 1.

Figure 9 shows the energy spectra at the 9 instants
(corresponding to different shock heights) as presented in Figure
8. The vertical coordinate is given by 
the number of energetic electrons ($\Delta$N) in a certain energy
range ($\Delta$($E/E_0$)). We can see all
these energy spectra can be well approximated by a power-law
spectra with an index of $\sim -3$. It is well known that the
DSA mechanism of energetic particle acceleration is
capable of producing power-law spectra while the SDA mechanism in the scattering-free limit not \citep{bell78,wu84}.
The reason we get a power-law spectrum via the SDA mechanism here is due to the fact that electrons repetitively
travel back to the shock and get accelerated because of the trapping effect of closed field lines
 and the weak pitch-angle scattering, physically equivalent to the diffusive process in DSA theory.
 Note similar power-law spectra have been
found in previous simulations within the framework of SDA mechanism \citep[e.g.,][]{burgess06,guo10}.

In addition, energetic electrons as modeled here are bi-directional.
Figure 10 shows the pitch angle distribution of energetic electrons ($E$ $>$5 $E_0$) in the shock upstream.
We can see that at most times it indicates a counter-streaming distribution.
Counter-streaming electrons may give rise to counter-propagating Langmuir waves and strong harmonic emission \citep{ganse12,ganse14}.
This, as well as the power-law distribution,
may be important to the excitation of enhanced Langmuir waves and type II bursts and should be considered in type II theories and simulations \citep[e.g.,][]{schmidt12a,schmidt12b,schmidt14}.

In the above text, we have demonstrated that the closed field lines play the role of an electron trap,
via which the electrons are sent back to the shock front multiple times,
and therefore accelerated to high energies by the shock.
We note that the present model is rather simplified and the case of a planar shock wider than the streamer and therefore the shock
 intersecting at two points with the same field line is a best case scenario. In the case of the shock having only one intersection with the streamer closed field, then the trapping will be realized by the shock and the magnetic mirror at the other end of the tube, the acceleration time may become much longer.
It should be further extended
by a parameter study on the effect of the streamer configuration,
the shock geometry, compression ratio, as well as the scattering
effect by coronal waves and turbulence. In addition, a more
realistic streamer configuration with a self-consistent
eruption-generated magnetohydrodynamic shock, like those simulated
by \citet{roussev04} and \citet{chen07}, should be adopted.

\section{Conclusions and Discussion}\label{sec4}

In this paper we first examine two coronal type II events
occurring on 2011 March 25 and 2011 March 27 and present their
common observational features (as follows). (1) Both events erupted
from the same AR below a streamer structure, and the sweeping
process of the CME front across the streamer structure can be well
observed; (2) In both events, the heights of the CME fronts as
measured with the coronagraph data are consistent with that
deduced from the type II spectral fittings; (3) Both type II
bursts ended once the CME fronts passed the streamer cusp, subject
to observational uncertainty. These observations, especially the last point,
lead us to conjecture that the closed field topology of the streamer may be important
to the shock acceleration of the type-II-emitting energetic electrons.

To validate this conjecture, we perform a test-particle simulation of electron acceleration in a shock-streamer system.
Simulation results show that only those electrons that are injected within the closed field regions can be accelerated
efficiently. The trapping effect of the closed streamer structure allows the electrons to return to the shock front multiple times and be
repetitively accelerated via the SDA mechanism. It is shown that energetic electrons mainly concentrate around the tips of relevant closed
field lines in the shock upstream, and are almost counter-streaming with energy spectra approximated by a power-law with an index of $\sim -3$. This predicts some features of the energetic electrons and the possible source location of the type II radio bursts, and provides an explanation
to the well-known narrowband feature of type II bursts. The simulation forms a basis
 for further studies with a more realistic configuration.

Our study highlights the possible role of large-scale magnetically closed
structures, as a trapping agency of energetic electrons, in shock-induced electron acceleration and metric type II excitation.
It is well known that a majority of solar eruptions
originate from closed field regions above the AR. If the
eruption-induced shock (either the flare blast wave or the CME-driven shock)
is formed inside the streamer, a
streamer-shock configuration as described here can arise.
Obviously the generation of such a configuration is not limited to
streamers. For any coronal loops or magnetic arcades, if the shock
is excited within them, an equivalent particle acceleration system
can present. Therefore, electron acceleration by a shock
propagating in a closed field structure may be important to the
generation of metric type IIs in more events.
Note that the scenario should be considered as complementary to any shock acceleration
mechanism. We do not intend to reject any other theories or
processes of shock-electron acceleration, e.g., those considering
the effect of shock ripples \citep{guo10}, turbulence
\citep{guo10}, and whistler waves \citep{oka06,wilson12}.

The relationship between coronal/metric type IIs and
interplanetary (IP) type IIs is a long-standing problem with
debate. As observed in radio dynamic spectra, metric and IP type
IIs often do not join with each other, and many metric type IIs do
not have IP counterparts \citep[e.g.,][]{gopalswamy01,cane05}. At
present, two scenarios are proposed for such observations. The
first scenario proposes that the Alfv\'enic speed maximum, which
is reached at $\sim$3-4 $R_\odot$, is responsible for the type II
discontinuity since the shock gets weakened or even disappeared at
this Alfv\'en maximum \citep{mann99,mann03,gopalswamy01,vrsnak02}.
The other scenario suggests that the metric and IP type IIs are
produced by shocks of different origins. The IP type IIs are
generated by the interplanetary CME-driven shock, while the
coronal type IIs may be from the flare-driven blast wave
\citep[e.g.,][]{gopalswamy98,cane05}. However, the latter
suggestion regarding the metric type II origin still suffers from
hot debate (see the recent review by \citet{vrsnak08}).

In this study, we provide an alternative explanation.
We suggest that there are metric type IIs that are closely related to the
closed field structures (e.g., streamers and coronal loops). Once
the shock goes beyond the outermost part of the closed field
structures, the metric type IIs may stop accordingly.
Observationally, it is known that many coronal type IIs end at
frequencies above $\sim$20-30 MHz
\citep[e.g.,][]{nelson85,shanmugaraju03}. These termination
frequencies correspond to the heliocentric coronal heights of 2-3
$R_\odot$ according to broadly-used density models
\citep[e.g.,][]{newkirk61,saito70}. On the other hand, it is
generally believed that coronal magnetic field structures are
closed below this distance. For IP type IIs and the rare type II
events that can extend from metric to decameter/hectometric
wavelength \citep[e.g.,][]{cane05}, their generations are
certainly not completely determined by the large-scale closed
coronal structures, and should be explained differently \citep[e.g.,][]{bale99,pulupa08}.

The simulation of Schmidt \& Cairns (2014) and relevant studies of the same group of authors represent the state-of-the-art study in type II theory and simulation. Their simulation is able to explain some important observational features of type IIs, including the type II frequencies, emission intensities, the overall drift rate and the spectral intermittencies. They also make some predictions about the radio emitting locations. For example, the sources may be away from the shock nose and may change with time as $\theta_{Bn}$ changes), and the emission intensities are greatly affected by the interaction of CMEs with coronal and interplanetary structures. In comparison, we notice that their simulation predicts a weak emission at the shock-streamer interaction region due to the change of $\theta_{Bn}$ there. This seems to be inconsistent with some observations that the CME-streamer interaction region is important to type II bursts (see our introduction). In the Schmidt \& Cairns study, the type II emissivity is mainly determined by the shock speed (strength) and the shock geometry (the value of $\theta_{Bn}$). The trapping effect of closed field lines on shock electron acceleration, as well as the effect of counter-streaming electrons on type II emissivity have not been considered. This might be the reason of the above inconsistency.


\acknowledgements

We are grateful to the STEREO, SDO, BIRS, and Learmonth teams for
making their data available to us. This work was supported by
grants NSBRSF 2012CB825601, NNSFC 41274175 and 41331068. Gang Li's
work at UAHuntsivlle was supported by NSF grants ATM-0847719 and
AGS1135432.

\begin{figure}
\includegraphics[width=0.95\textwidth]{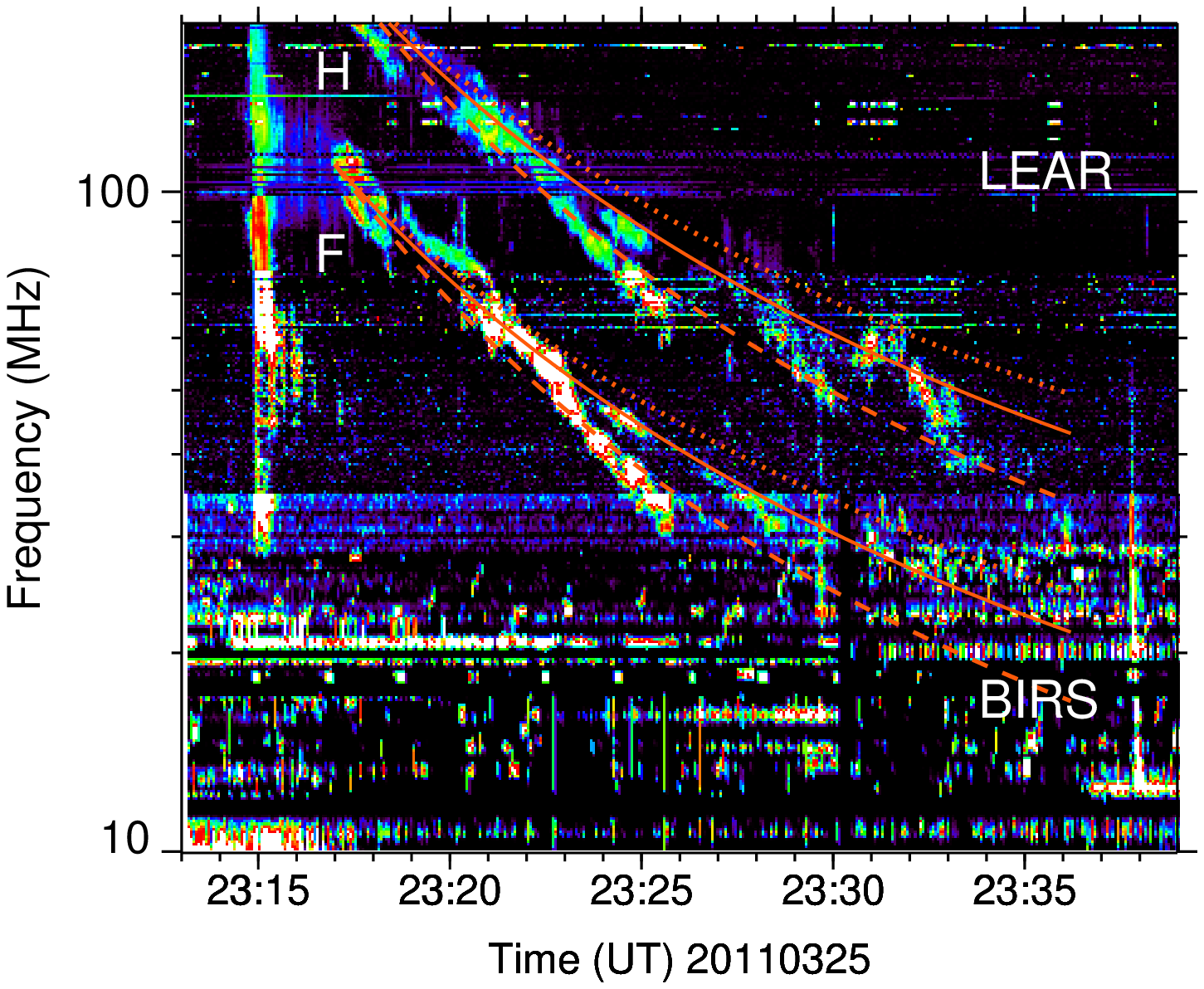}
\caption{The radio dynamic spectrum on 2011 March 25 as recorded
by BIRS (10-35 MHz) and Learmonth (35-180 MHz) radio
spectrometers. ``F'' and ``H'' denote the fundamental and harmonic
bands of the type II burst. The red lines are the fitting curves
using 1-fold (dashed), 2-fold (solid) and 3-fold (dotted) Newkirk
density model and a shock speed of 620 km s$^{-1}$,
respectively.}\label{Fig1}
\end{figure}

\begin{figure}
\includegraphics[width=0.9\textwidth]{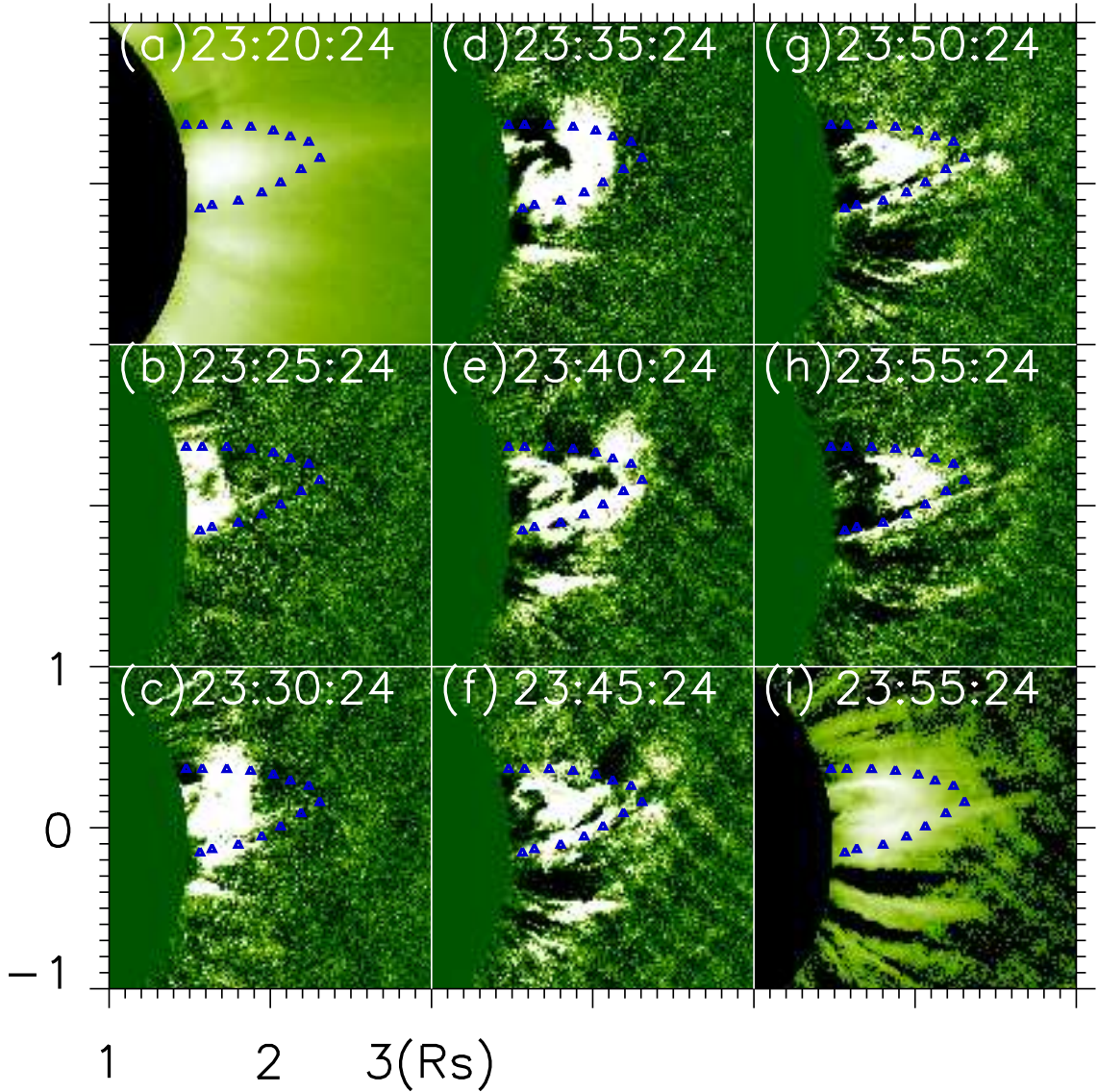}
\caption{STEREO/COR1 B white light data at
23:20-23:55 UT on 2011 March 25. Panel (a) is the direct image, panels (b)-(h)
are running difference images, and panel (i) is the base difference image (23:55 UT - 23:20 UT).
The blue triangles are the outlining streamer envelop depicted from panel (a) and over-plotted onto other panels to indicate the relative
location of the streamer and the CME front.
}\label{Fig2}
\end{figure}

\begin{figure}
\includegraphics[width=0.75\textwidth]{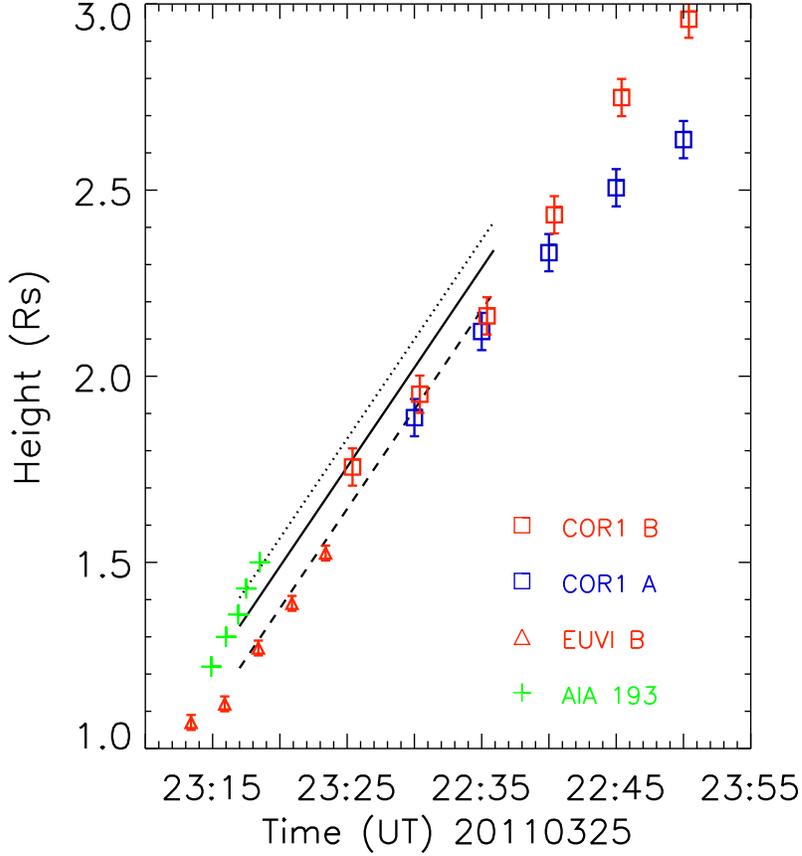}
\caption{The heliocentric heights of CME
fronts measured from COR1 A (blue squares), COR1 B (red squares)
and EUVI B (red triangles). The error bars indicate measurement
uncertainties of the heights, estimated to be $\sim$0.05 $R_\odot$
for COR1 and $\sim$0.02 $R_\odot$ for EUVI. The green ``+'' signs
are heights of coronal EUV wave observed by AIA in 193 \AA\ (taken
from \citet{kumar13}). The heights of the radio source deduced from
the type II spectral fittings (see Figure 1) are shown as three
black straight lines.}\label{Fig3}
\end{figure}

\begin{figure}
\includegraphics[width=0.6\textwidth]{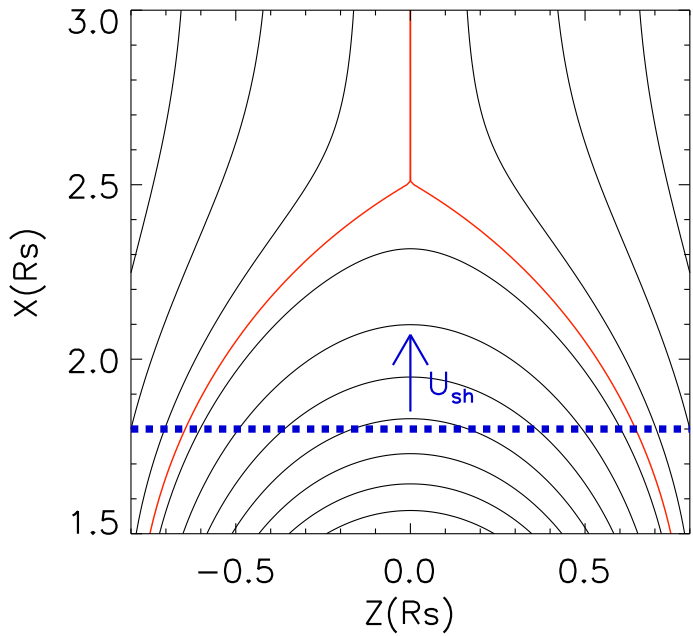}
\caption{The streamer-shock configuration in the numerical
simulation. The streamer magnetic field is given by an analytical
model in \citet{low86}. The black lines represent magnetic field
lines, the line in red presents the outermost closed field line
and the current sheet above, and the height of streamer cusp is
taken to be 2.5 $R_\odot$. A planar shock front propagating along
the streamer axis is shown by the blue dashed line.} \label{Fig4}
\end{figure}

\begin{figure}
\includegraphics[width=0.7\textwidth]{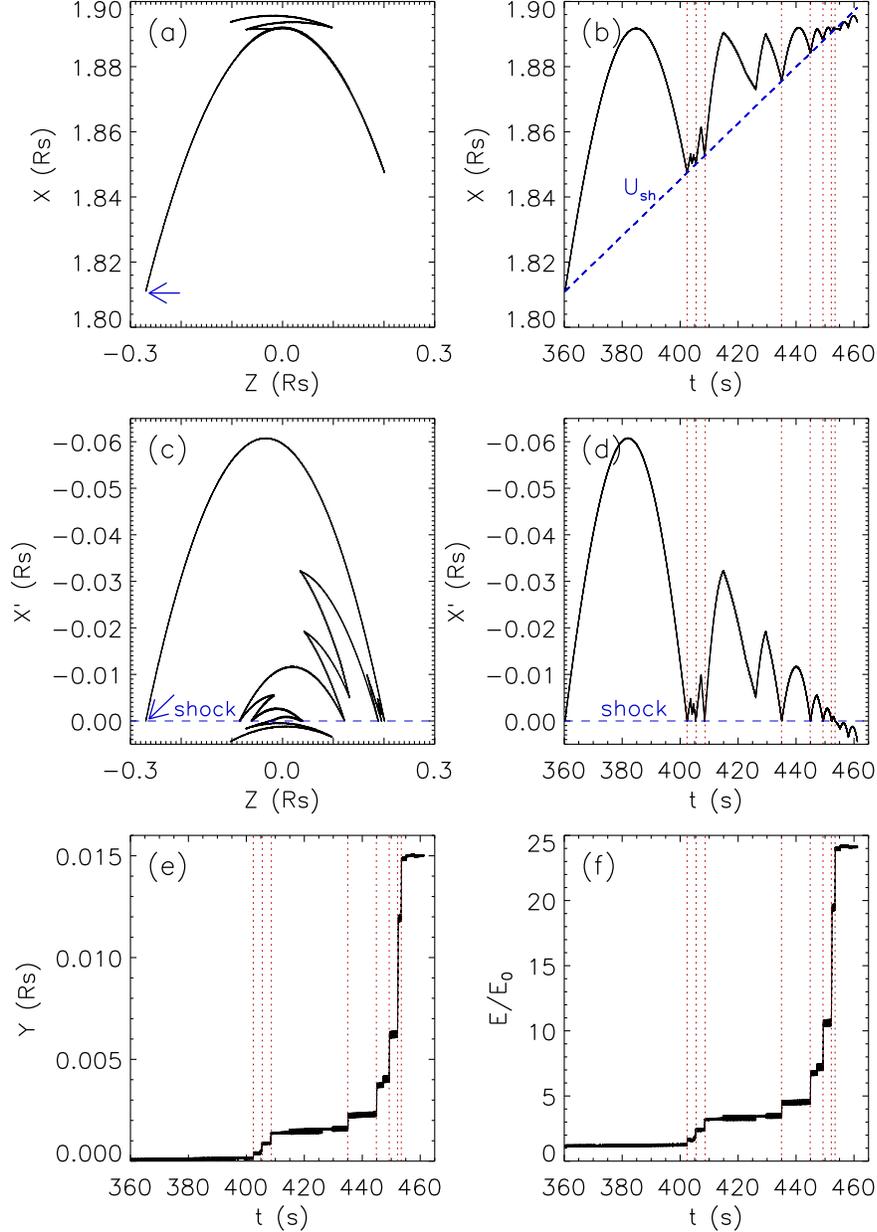}
\caption{Simulation results for an electron that is accelerated to $\sim$25 $E_0$. Panels (a) and (c)
display the electron trajectory in the $x$-$z$ lab frame and in
the $x'$-$z$ shock frame. The blue arrows in the two panels point
to the injection point of the electron. Panels (b) and (d) show
its position $x$ and $x'$ over time, respectively. The dashed blue
line in panel (b) indicates the position of the outward
propagating shock, while that in panels (c) and (d) represents the
shock front ($x'$ = 0) in the shock frame. Panels (e)-(f) present
the temporal evolution of the electron drifting distance along the
$y$ direction and the temporal evolution of its energy in the
shock frame. The vertical red dotted lines denote the electron
reflection points at the shock.} \label{Fig5}
\end{figure}

\begin{figure}
\includegraphics[width=0.95\textwidth]{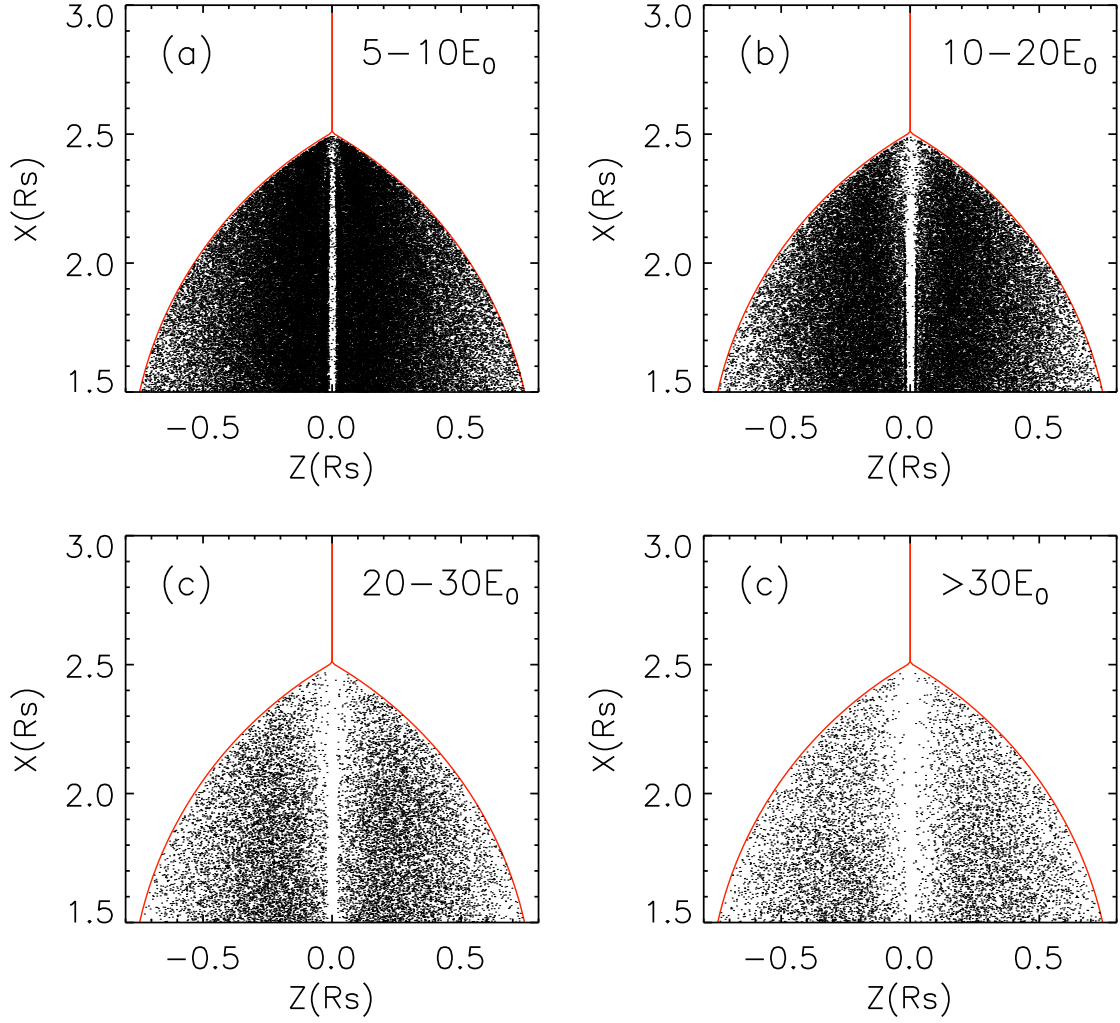}
\caption{The distribution of injection position of electrons that
have achieved a final energy of 5-10 $E_0$, 10-20 $E_0$, 20-30
$E_0$ and $>$30 $E_0$, respectively. } \label{Fig6}
\end{figure}

\begin{figure}
\includegraphics[width=0.6\textwidth]{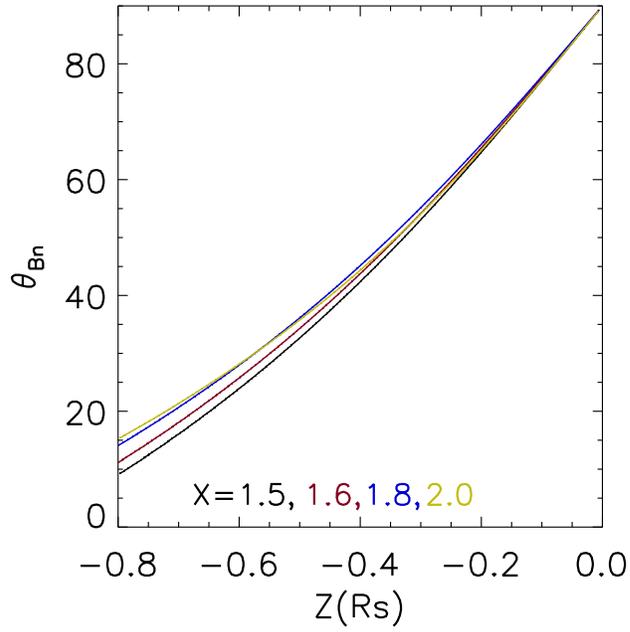}
\caption{Variations of $\theta_{Bn}$ (the angle between the
upstream magnetic field line and the shock normal) with $z$ at
different distances ($x$ = 1.5, 1.6, 1.8 and 2.0 $R_\odot$).}
\label{Fig7}
\end{figure}

\begin{figure}
\includegraphics[width=0.95\textwidth]{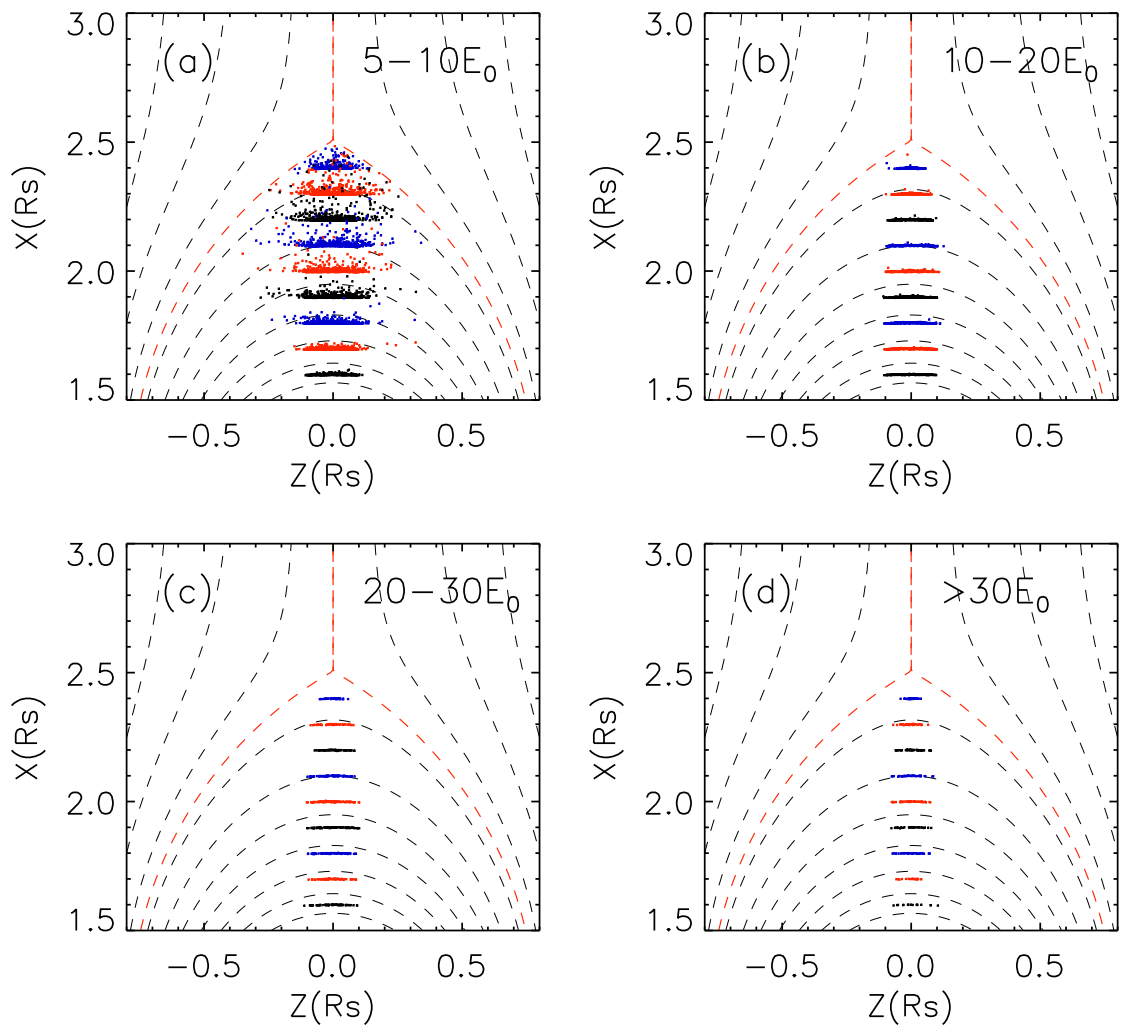}
\caption{The electron positions at several different times when
the shock propagates to distances of 1.6, 1.7, 1.8, ..., 2.4
$R_\odot$. These 9 different data sets are superposed onto the
streamer field lines, where the scattering points in panels
(a)-(d) represent electrons that have been accelerated to 5-10
$E_0$, 10-20 $E_0$, 20-30 $E_0$, and $>$30 $E_0$, respectively.}
\label{Fig8}
\end{figure}

\begin{figure}
\includegraphics[width=0.95\textwidth]{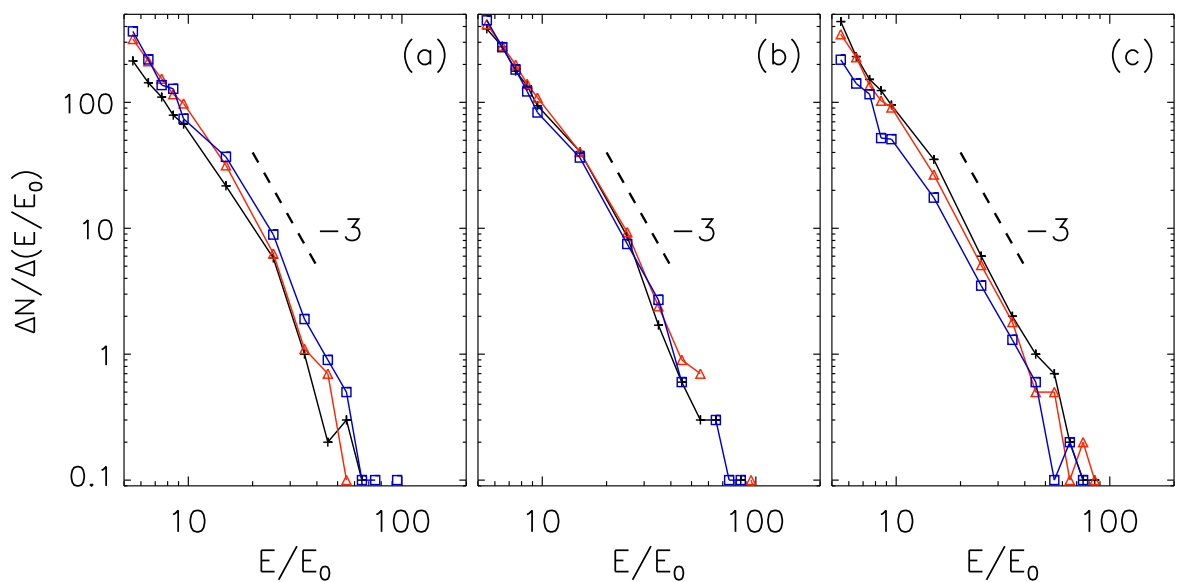}
\caption{The electron energy spectra at the 9 different times
(corresponding to different shock heights $x$ =1.6, 1.7, 1.8, ..., 2.4
$R_\odot$) as presented in Figure
8, shown in each panel as black pluses, red triangles and blue
squares with increasing shock heights. In panel (a): $x$ = 1.6,
1.7 1.8 $R_\odot$; panel (b): $x$ = 1.9, 2.0, 2.1 $R_\odot$; panel
(c): $x$ = 2.2, 2.3, 2.4 $R_\odot$. The black dashed line
represents a -3 power-law spectrum.} \label{Fig9}
\end{figure}

\begin{figure}
\includegraphics[width=0.95\textwidth]{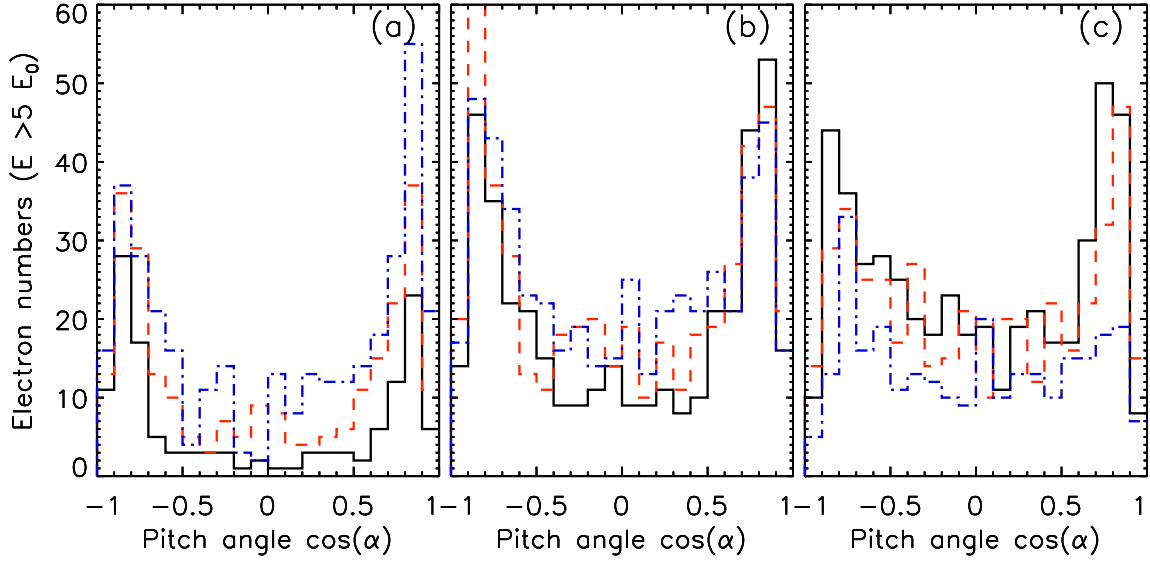}
\caption{The electron pitch angle at the 9 different times
(corresponding to different shock heights  $x$ =1.6, 1.7, 1.8, ..., 2.4
$R_\odot$) as presented in Figure
8, shown in each panel as black solid lines, red dash line and blue
dash-dot line with increasing shock heights. In panel (a): $x$ = 1.6,
1.7 1.8 $R_\odot$; panel (b): $x$ = 1.9, 2.0, 2.1 $R_\odot$; panel
(c): $x$ = 2.2, 2.3, 2.4 $R_\odot$. } \label{Fig10}
\end{figure}

\end{document}